\documentclass[%
showkeys,
amsmath,amssymb, aps, prc, nofootinbib, twocolumn
]{revtex4-1}

\usepackage{graphicx}
\usepackage{dcolumn}
\usepackage{multirow}
\usepackage{bm}

\begin{document}

\title{Collective flow at SIS energies within a hadronic transport approach: Influence of light nuclei formation and equation of state}
\author{J.-Mohs$^{1,2}$, M.~Ege$^{2}$, H.~Elfner$^{3,2,1}$ and M.~Mayer$^{2}$}
\affiliation{$^1$Frankfurt Institute for Advanced Studies, Ruth-Moufang-Strasse 1, 60438
Frankfurt am Main, Germany}
\affiliation{$^2$Institute for Theoretical Physics, Goethe University,
Max-von-Laue-Strasse 1, 60438 Frankfurt am Main, Germany}
\affiliation{$^3$GSI Helmholtzzentrum f\"ur Schwerionenforschung, Planckstr. 1, 64291
Darmstadt, Germany}
\date{\today}

\begin{abstract}
Collective flow observables are known to be a sensitive tool to gain insights on the equation of state of nuclear matter from heavy-ion collision observations. Towards more quantitative constraints one has to carefully assess other influences on the collective behaviour. In this work a hadronic transport approach SMASH (Simulating Many Accelerated Strongly-interacting Hadrons) is applied to study the first four anisotropic flow coefficients in Au+Au collisions at $E_{\rm lab}=1.23A$ GeV in the context of the recently measured data by the HADES collaboration. In particular, the formation of light nuclei is important in this energy regime. Two different approaches are contrasted to each other: A clustering algorithm inspired by coalescence as well as microscopic formation of deuterons via explicit cross-sections. The sensitivity of directed and elliptic flow observables to the strength of the Skyrme mean field is explored. In addition, it is demonstrated that the rapidity-odd $v_3$ coefficient is practically zero in this energy regime and the ratio of $v_4/v_2^2$ is close to the value of 0.5 expected from hydrodynamic behaviour. This study establishes the current understanding of collective behaviour within the SMASH approach and lays the ground for future more quantitative constraints on the equation of state of nuclear matter within improved mean field calculations. 
\end{abstract}

\maketitle

\section{Introduction}

In heavy-ion collisions the anisotropy of the particle production in the transverse plane is described in terms of flow coefficients.
In transport calculations flow observables were shown to be very sensitive to nuclear potentials and the equation of state (EoS) \cite{Scheid:1974zz,Danielewicz:2002pu}.
Hence one can put constraints on the EoS by comparing hadronic transport calculations with experimental flow data.

Dependent on the active degrees of freedom in the different transport approaches different conclusions about the strength of the mean-field interactions are drawn. Approaches mainly based on nucleons and pions with sophisticated potentials including momentum dependent interactions as described in \cite{Gale:1987zz} come to the best agreement with the wealth of existing FOPI data \cite{FOPI:2011aa}, when mean fields corresponding to a 'soft' equation of state are employed \cite{Aichelin:1987ti, Isse:2005nk,Fevre:2015fza,LeFevre:2016vpp}. On the other hand, approaches like UrQMD with many resonance states arrive at the same quality of agreement with experimental data incorporating a mean field corresponding to a 'hard' equation of state \cite{Petersen:2006vm, Hillmann:2018nmd,Hillmann:2019wlt}.
Modifying the equation of state via modifications of the collision term offers an orthogonal avenue to study collective behaviour within transport approaches \cite{Nara:2016hbg}.

The nuclear equation of state at high densities is relevant also for the dynamical description of neutron star mergers. After the first detection of gravitational wave signals \cite{Abbott:2016blz}, there is a large interest in sophisticated theoretical calculations of their detailed evolution which is relevant for the nucleosynthesis including different assumptions about the equation of state of QCD matter at high densities (see \cite{Bauswein:2018bma,Most:2018hfd} for recent examples). Despite the isospin difference that is encoded in the symmetry energy that can also be extracted from heavy-ion reactions \cite{Wang:2020dru, Tsang:2012se}, further knowledge on the equation of state is important. Before a quantitative extraction including uncertainty quantification using Bayesian techniques is sensible, all possible systematic uncertainties in the transport approaches have to be understood. One of those is the formation of light clusters that incorporates about half of the nucleons in heavy-ion reactions at few GeV per nucleon beam energy. Coalescence models based on phase-space separation \cite{Butler:1961pr,Schwarzschild:1963zz,Gutbrod:1988gt} have been devised \cite{Sombun:2018yqh} as well as more comprehensive dynamical approaches \cite{Oh:2009gx, Aichelin:2019tnk}. Using coalescence, protons and deuterons are predicted to follow a constituent number scaling \cite{Hillmann:2019wlt}.

In this work we assess how SMASH-1.8 (Simulating Many Accelerated Strongly-interacting Hadrons) \cite{Weil:2016zrk, dmytro_oliinychenko_2020_3742965} performs compared to recent measurements of flow coefficients at SIS-18 enegies \cite{Kardan:2018hna, Adamczewski-Musch:2020iio} with a rather simple Skyrme parametrization of the EoS \cite{Xu:2016lue} and investigate the sensitivity of the results to the stiffness of the EoS. This is meant as a baseline for further calculations employing more sophisticated mean field implementations such as the one in \cite{Sorensen:2020ygf}. The main emphasis in this work is on exploring different treatments for light nuclei production, which constitutes one of the major uncertainties. We concentrate on the most abundant lightest one namely the deuteron and compare two different ways of taking the light nuclei formation into account. Coalescence of nucleons in the final state as described in \cite{Sombun:2018yqh} is contrasted to the dynamical treatment by producing and propagating deuterons throughout the evolution of a heavy-ion collision as described in \cite{Oliinychenko:2018ugs}.

The manuscript is structured as follows:
First the SMASH transport with the current state of nuclear potentials present in the calculation and the different ways of treating light nuclei formation is described in Section \ref{sec_theory}.
Then the effect of different Skyrme parameters and light nuclei treatments on the directed flow in Section \ref{sec_directed_flow} and elliptic flow in Section \ref{sec_elliptic_flow} of nucleons and deuterons are evaluated.
Afterwards the evolution of  the directed and elliptic flow with time is studied in Section \ref{sec_evolution}. Last we present some higher flow coefficients and show the $v_3$ computed with the scalar product method and the ratio of $v_4/v_2^2$ in Section \ref{sec_higher_coefficients}.
Finally we conclude and present an outlook in Section \ref{sec_conclusions}.

\section{SMASH with potentials and light nuclei production}
\label{sec_theory}
\subsection{Hadronic transport with mean field}
\label{sec_mean_fields}

In this work, we apply the transport approach SMASH version 1.8 \cite{Weil:2016zrk, dmytro_oliinychenko_2020_3742965} to calculate flow observables.
The relativistic Boltzmann equation is effectively solved in a testparticle ansatz with hadronic degrees of freedom.
Starting with nulcei sampled from a Woods-Saxon distribution hadrons are explicitly propagated between interactions.
At low collision energies most inelastic collisions form resonances with properties adopted from \cite{Tanabashi:2018oca}.
For the resonances the spectral functions are described by relativistic Breit-Wigner distributions without medium modification.
The mass dependence of the decay width is included according to \cite{Manley:1992yb}.

For the propagation between the interactions nuclear potentials are optionally taken into account. Let us note here, that for the current work we stick with a simplified treatment, since this is the same setup that has been used in other recent works in comparable approaches \cite{Hillmann:2018nmd} and the default parameters have been determined in a comparison between multiple different transport codes \cite{Xu:2016lue}. 

At this point the Skyrme and symmetry potential are present in the calculation
\begin{equation}
	U=U_{\rm Sk} + U_{\rm Sym}.
\end{equation} 
The contributions to the potential energy is given in terms of the density as
\begin{equation}
	U_{\rm Sk} = A\left(\frac{\rho_B}{\rho_0}\right) + B\left(\frac{\rho_B}{\rho_0}\right)^\tau,
\end{equation}
where $\rho_B$ is the net-baryon density, $\rho_0 = 0.168\,\rm fm^{-3}$ is the nuclear ground state density and $A$, $B$ and $\tau$ are parameters.
Furthermore
\begin{equation}
	U_{\rm Sym} = \pm 2S_{\rm pot}\frac{\rho_{I_3}}{\rho_0},
\end{equation}
where $\rho_{I_3}$ is the density of the relative isospin projection $I_3/I$, the sign is positive for positive isospin and negative for negative isospin of the particle of interest.

Since the potentials are not written in  a covariant form, they have to be evaluated in the local rest frame for a Lorentz invariant treatment.
In practice, the calculation is performed in an arbitrary calculation frame such as the fixed target or center of mass frame. 
Therefore a boost needs to be introduced which results in the following form for the force acting on a particle in the calculation frame:
\begin{equation}
\begin{split}
\vec{F} &= \frac{\partial U_\mathrm{Sk}}{\partial \rho_B} \left[-\left( \vec{\nabla} \rho_B + \partial_t \vec{j}_B\right) + \dot{\vec{x}} \times\left(\vec{\nabla} \times \vec{j}_B\right) \right] \\
&+ \frac{\partial U_\mathrm{Sym}}{\partial \rho_{I_3}} \left[-\left( \vec{\nabla} \rho_{I_3} + \partial_t \vec{j}_{I_3}\right) + \dot{\vec{x}} \times\left(\vec{\nabla} \times \vec{j}_{I_3}\right) \right]
\end{split}
\end{equation}
$\vec{j}_B$ and $\vec{j}_{I_3}$ are the net-baryon current and the $I_3$ current respectively and $\dot{\vec{x}}$ is the velocity of the particle of interest.
Given the force in the calculation frame, the spacial components $p_i$ of the particle momentum are updated at each time step
\begin{equation}
	p_i \rightarrow p_i + F_i\Delta t\,.
\end{equation} 
The calculation therefore relies on small time steps.

Since the potentials and the equations of motion depend on the densities and their spacial derivatives, a high resolution in the density is necessary.
In the calculations with potentials within this work, each particle is represented by 20 test-particles for a more precise estimate of the densities and their derivatives.  

Different sets for the parameters of the Skyrme potential are described in \cite{Kruse:1985hy}.
They are labeled soft and hard corresponding to their stiffness.
In addition, we include the default parameter set in SMASH \cite{Xu:2016lue}, which lies between the soft and hard one in terms of stiffness.
The parameter sets are explicitly given in Table \ref{tab_parameters}.

In the future one should aim for a more continuous variation of parameters to find the best values based on experimental data within a Bayesian multi-parameter analysis. 
The momentum dependence will be added and the implementation of relativistic mean fields based on density functional theory is work in progress \cite{Sorensen:2020ygf}. In the current work, we do not aim yet at a quantitative extraction of the nuclear equation of state, but would like to provide a baseline for further studies and investigate the influence of light cluster formation.

\begin{table}
	\begin{tabular}{|c|c|c|c|}
		\hline
		           & soft & default & hard \\
		           \hline
		$A$    & $-365$ MeV & $-209.2$ MeV & $-124$ MeV \\
		$B$      & $303$ MeV & $156.4$ MeV & $71$ MeV \\
		$\tau$ & $1.17$ & $1.35$ & $2.0$ \\
		$K$ & $200$ MeV & $240$ MeV& $375$ MeV\\
		\hline
	\end{tabular}
	\caption{Parameter sets for Skyrme potential with corresponding compressibilities $K$.}
	\label{tab_parameters}
\end{table}
  
\subsection{Clustering and light nuclei}
\label{sec_clustering}
At low beam energies, many protons are bound into light nuclei \cite{Reisdorf:2010aa}.
Therefore light nuclei formation needs to be taken into account from the theory side to be comparable with experiments.
One option to do so is to employ a coalescence model.
Here one can directly calculate the spectrum of nuclei given the proton spectra \cite{Butler:1961pr,Schwarzschild:1963zz,Gutbrod:1988gt} by assuming nucleons to form clusters once they are sufficiently close in phase space.
A similar approach is to perform clustering on a microscopic basis in each event separately also using the distance in phase space as a criterion for coalescence (see \cite{Sombun:2018yqh}).
Finally, one can explicitly form nuclei through production cross sections from hadrons and treat the nuclei dynamically as active degrees of freedom as described in Section \ref{sec_deuterons}.

In the default SMASH setup, light nuclei are not treated explicitly as degrees of freedom.
Instead they are made up of individual nucleons that interact via nuclear potentials.
Despite the potentials, nucleons do not form stable bound states in the final state of a heavy ion collision.
We therefore follow the method of identifying light nuclei as described in \cite{Sombun:2018yqh}, where the condition for two particles to form a cluster is a small distance in both momentum and coordinate space in the center of mass frame of the two particles at the time where the last interaction occurs in which one of the two particles took part. 
Compared to the simple clustering algorithm used in \cite{Weil:2016zrk} the main difference is the time at which the coordinate space distance of particles is evaluated.
The threshold for the momentum and position distances cannot be adopted from \cite{Hillmann:2019wlt}, where the improved method was used for a similar study, since the particles are not represented by multiple test-particles in that work and the average distance in phase-space is therefore not comparable.
Instead, new thresholds $r_0$ and $p_0$ in coordinate and momentum space are found based on the proton rapidity spectrum at $1.23A$ GeV using Bayesian parameter estimation where the values $r_0=0.87^{+0.03}_{-0.03}$ fm and $p_0 = 0.43^{+0.03}_{-0.02}$ GeV were found.

\subsection{Deuterons in SMASH}
\label{sec_deuterons}
A second option to consider the formation of light nuclei was introduced in \cite{Oliinychenko:2018ugs} and successfully applied to describe the centrality dependence of deuteron formation at LHC energies \cite{Oliinychenko:2020ply} as well as the energy dependence \cite{Oliinychenko:2020znl}.
The idea is to explicitly implement cross sections for nuclei production and treat the nuclei as active degrees of freedom as previously done in \cite{Danielewicz:1991dh,Oh:2009gx}.
The relevant processes for deuteron formation including protons $p$, neutrons $n$, pions $\pi$ and generic nucleons $N$ are
\begin{eqnarray}
	pnN & \leftrightarrow & dN \\
	pn\bar{N} & \leftrightarrow & d\bar{N} \\
	pn\pi & \leftrightarrow & d\pi \\
	NN & \leftrightarrow & d\pi .	
\end{eqnarray}
The first two are $3\leftrightarrow2$ processes that are in practice realized with $2\leftrightarrow1$ and $2\leftrightarrow2$ processes since multiparticle interactions are not yet present in SMASH.
For the intermediate step, a fictional dibaryon resonance $d' \leftrightarrow pn$ is used that can react in $2\leftrightarrow 2$ processes to a deuteron like $Nd'\leftrightarrow Nd$ or $\pi d'\leftrightarrow\pi d$.

\begin{figure}
	\centering
	\includegraphics[width=0.5\textwidth]{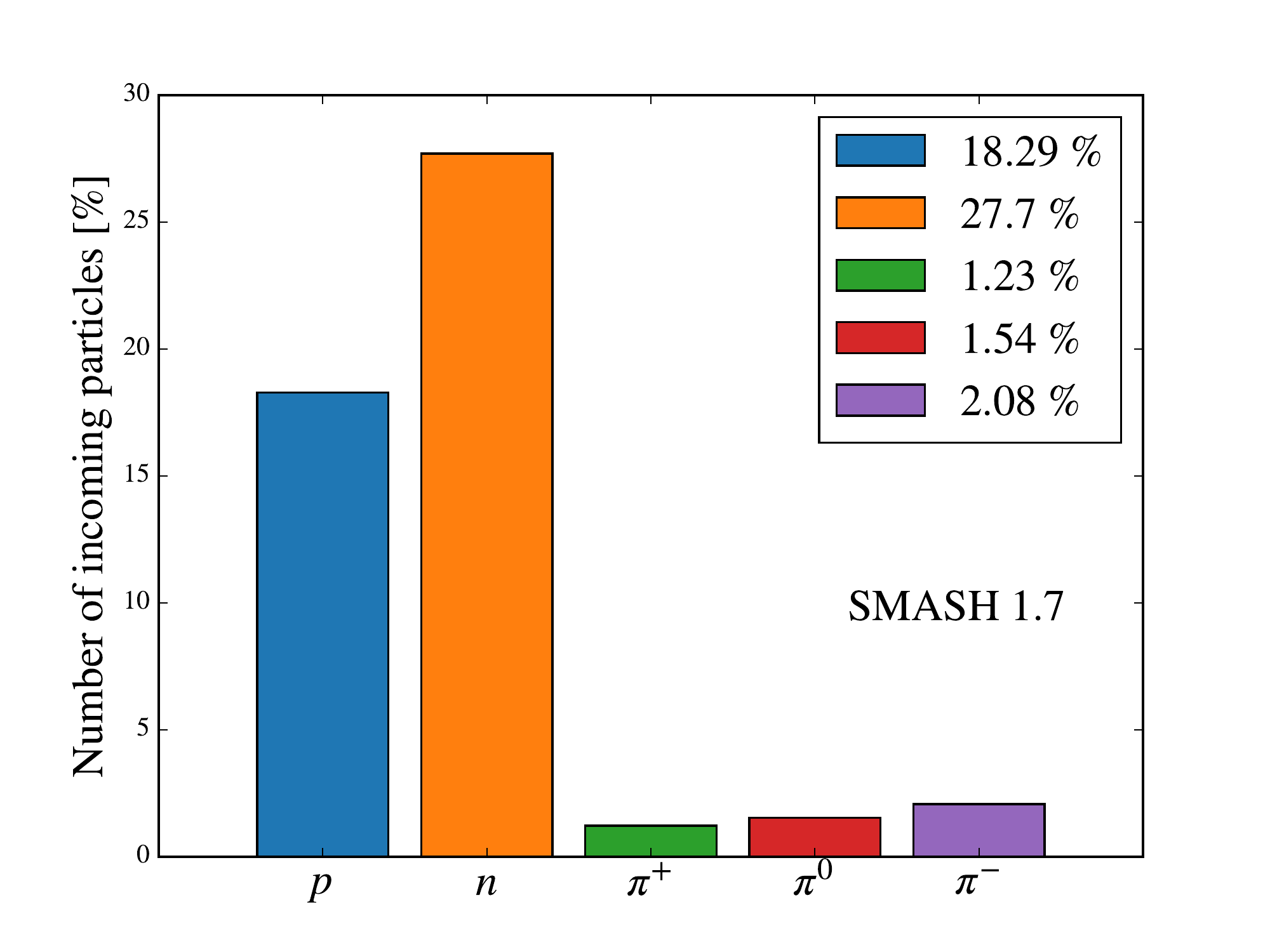}
	\caption{Relative number of reaction partners of the fictional $d'$ resonance in gold-gold collisions at $E_{\rm kin}=1.23A$ GeV. The $d'$ often decays into $pn$ without a collision partner, therefore the numbers don't sum up to 100\%.}
	\label{fig_reactionpartners}
\end{figure}

To investigate the relevant processes for deuteron formation Figure \ref{fig_reactionpartners} shows the relative number of reaction partners of the $d'$ resonance in percent in gold-gold collisions at $E_{\rm kin} = 1.23A$ GeV. The numbers do not sum up to 100\% because in many cases the $d'$ decays without a collision partner.
The nucleons clearly dominate the reaction partners, so the $pnN\leftrightarrow dN$ process is as expected the most important at this energy \cite{Danielewicz:1991dh}.

At higher energies, the most important process is the one involving a pion \cite{Oliinychenko:2018ugs}.
This difference is caused by the fact that at higher energies more pions are produced while at low energies a dense medium of nuclear matter is produced.

In this work, we compare the simple coalescence approach with the explicit deuteron formation to investigate which description is more favored by the experimental data on flow measurements. 
The two approaches differ in the sense that the coalescence takes the formation of arbitrary clusters into account while at this point the explicit formation of only deuterons is implemented.
It is possible to extend the model also to larger nuclei but the complexity of the problem increases with the mass number.

\begin{figure}
	\centering
	\includegraphics[width=0.5\textwidth]{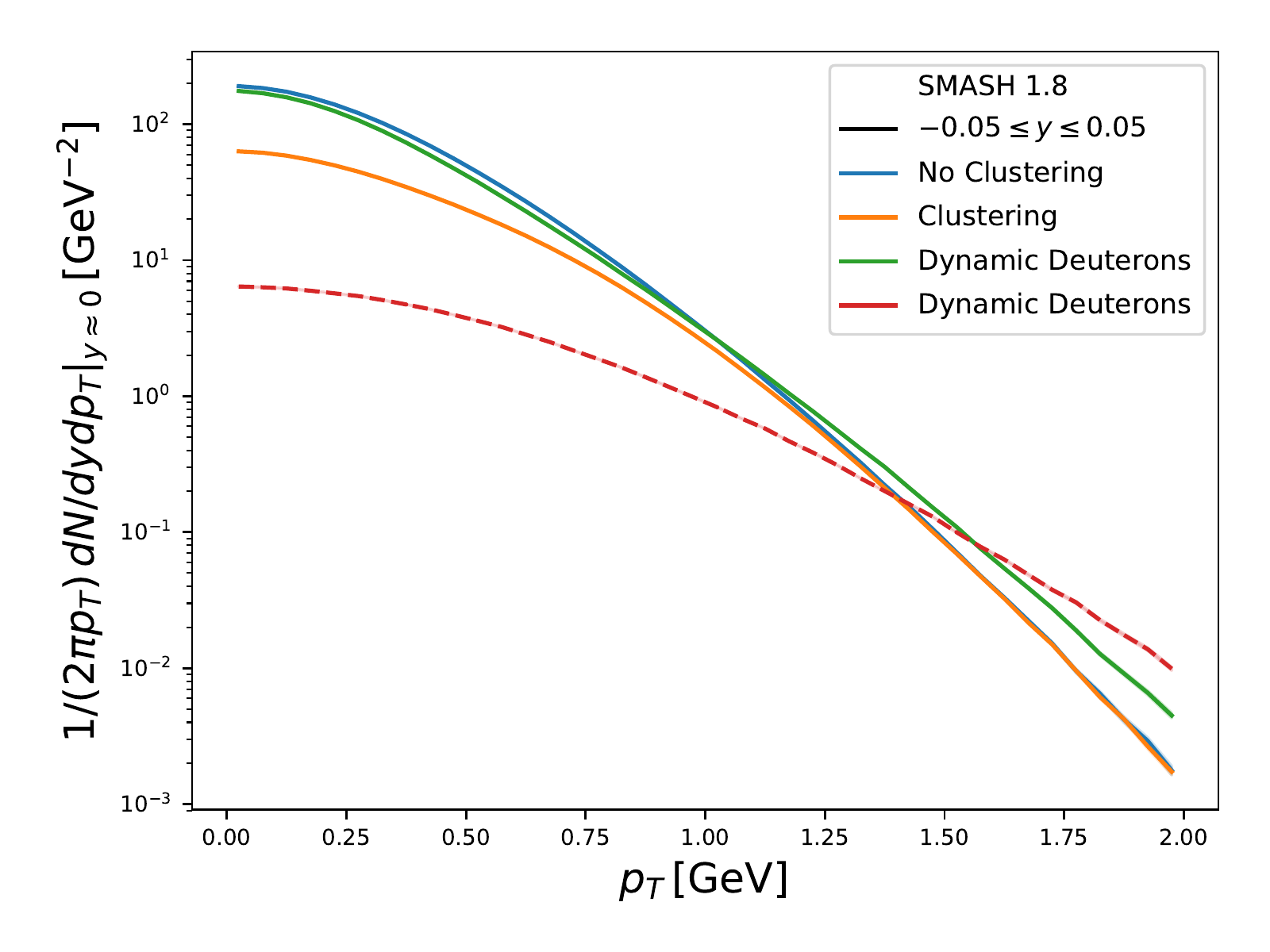}
	\caption{Transverse momentum spectra of nucleons (full lines) and deuterons (dashed lines) at mid-rapidity in 20\%-30\% central gold-gold collisions at $E_{\rm kin}=1.23A$ GeV. For all calculations the hard equation of state is used.}
	\label{fig_pt_spectra}
\end{figure}

In Figure \ref{fig_pt_spectra} a comparison of the two methods in terms of the $p_T$-spectra of nucleons is shown. The curve labeled "no clustering" refers to a calculation without dynamic deuterons and without clustering.
Naturally, the number of nucleons is the largest in this case.
The clustering result has the lowest number of nucleons because the formation of different light nuclei can be taken into account as compared to the dynamic deuteron calculation.
The clustering mainly affects the low momentum part of the spectrum, where the density in phase space is large. 
Comparing the dynamic treatment of deuterons to the other curves, one can see that the deuteron production influences the dynamics of nucleons significantly as nucleons are shifted to larger $p_T$. 
This is related to interactions with deuterons that also have large transverse momentum on average.

\subsection{Flow calculations}
In the following we present results for the flow coefficients of protons and deuterons and compare to experimental data \cite{Kardan:2018hna, Adamczewski-Musch:2020iio}.
To be consistent with the measurement, the flow coefficients are evaluated with respect to the reaction plane which is fixed to the x-z-plane in the calculation setup.
Hence, using the angle $\phi= \arctan(p_y/p_x)$ the $n$-th order flow coefficient can be evaluated as an average over particles
\begin{equation}
	v_n=\langle \cos(n\phi)\rangle .
\end{equation}
The centrality of the events is selected by constraining the impact parameter to a specific range. We focus on the 20\% to 30\% most central gold-gold collisions, which correspond to the impact parameter range $6.6\,\mathrm{fm} <b< 8.1\,\mathrm{fm}$ \cite{Adamczewski-Musch:2017sdk}.

\section{Directed flow of protons and deuterons}
\label{sec_directed_flow}
\begin{figure}
	\centering
	\includegraphics[width=0.45\textwidth]{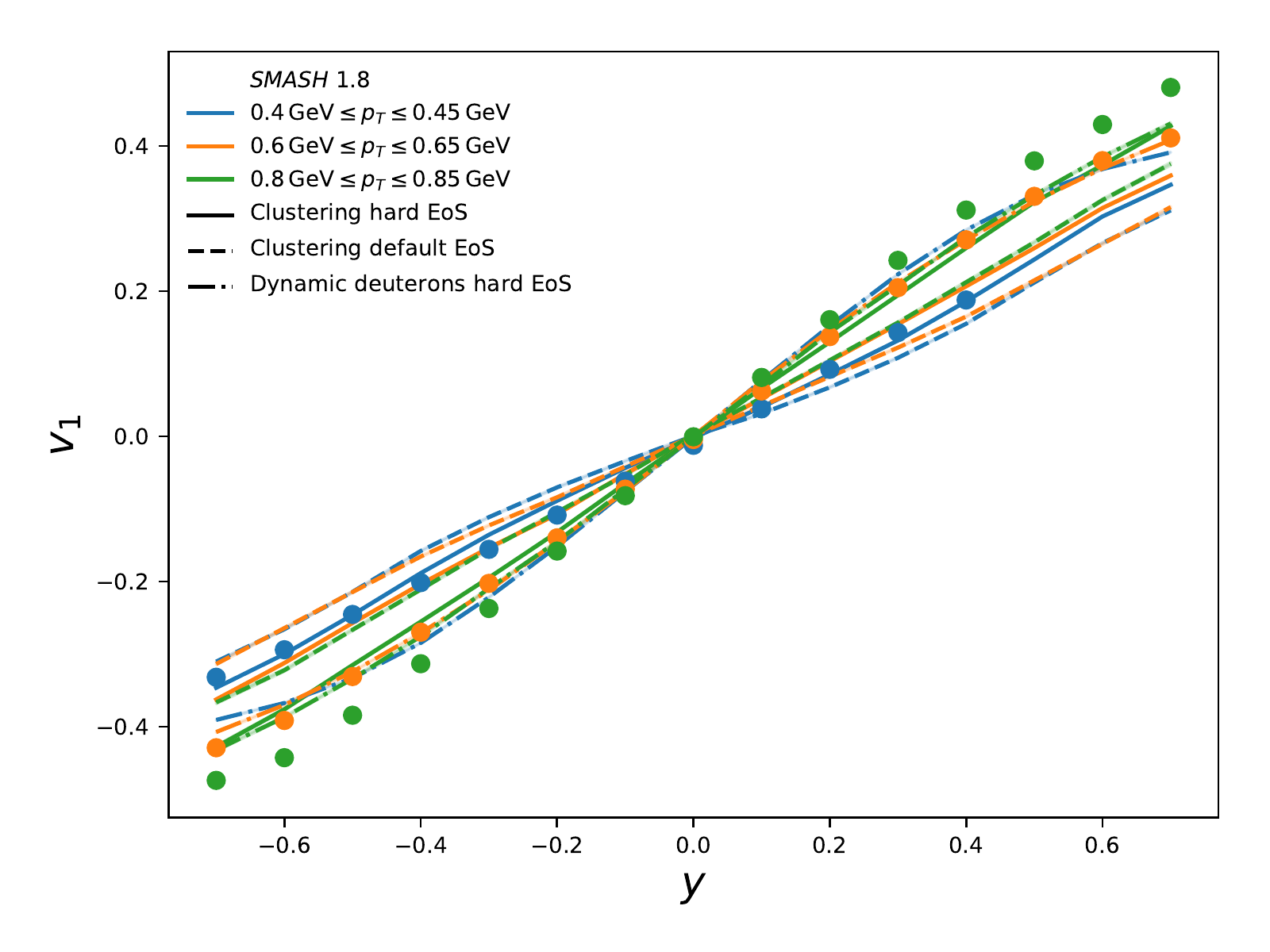}
	\caption{Directed flow of protons as a function of rapidity in 20\%-30\% central gold-gold collisions at $E_\mathrm{kin} = 1.23A\,\mathrm{GeV}$ for different $p_T$ bins compared to experimental data points \cite{Kardan:2018hna}. The full lines are obtained with a hard equation of state while for the dashed lines the default equation of state is used.}
	\label{fig_v1_y}
\end{figure}

We begin with the first order flow coefficient of protons and deuterons in gold-gold collisions at $E_{\rm kin} = 1.23A$ GeV.
The selection of rapidity and transverse momentum bins is chosen to cover a wide region in phase space. 
Since flow coefficients in general are sensitive to the strength of the potentials, results from calculations using different equations of state as introduced in Section \ref{sec_mean_fields}  are compared.
Figure \ref{fig_v1_y} shows the $v_1$ of nucleons as a function of rapidity for different bins in transverse momentum.
The shape reproduces the experimental measurements very well but only the calculation with the hard equation of state gives the correct magnitude.
Two calculations consider the formation of light nuclei by clustering in the final state as described in Section \ref{sec_clustering}.
In addition to the two equations of state, a third calculation is shown, where no clustering is performed but the deuterons are explicitly propagated as degrees of freedom in the calculation as described in Section \ref{sec_deuterons}.
For that calculation, also the hard equation of state is used.

\begin{figure}
   	\centering
   	\includegraphics[width=0.45\textwidth]{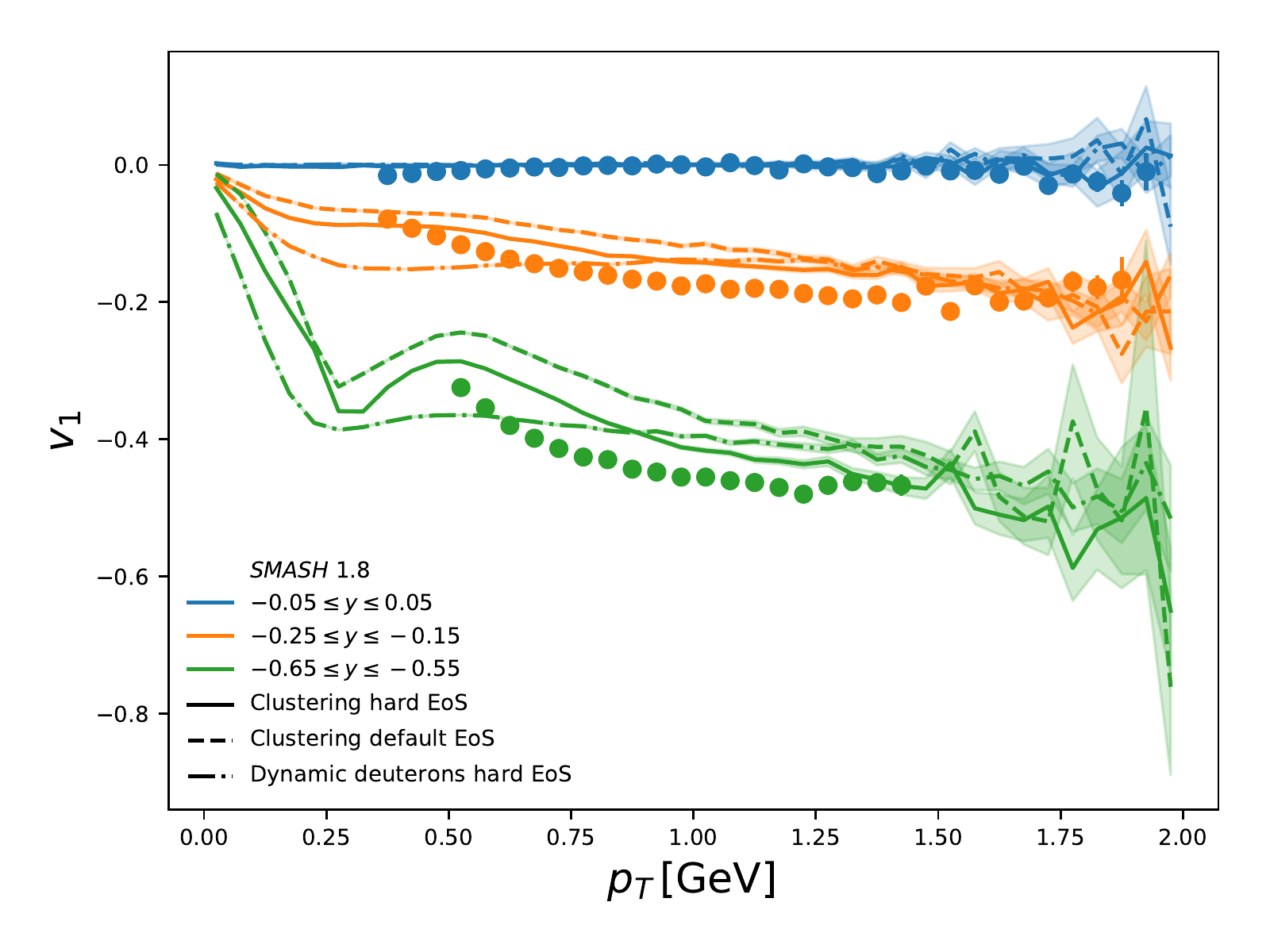}
   	\caption{Directed flow of protons as a function of transverse momentum in 20\%-30\% central gold-gold collisions at $E_\mathrm{kin} = 1.23A\,\mathrm{GeV}$ for different rapidity bins compared to experimental data points \cite{Kardan:2018hna}. The full lines are obtained with a hard equation of state while for the dashed lines the default equation of state is used. The curve labeled "No clustering" takes the formation of light nuclei into account by explicitly producing deuterons during the collision as described in Section \ref{sec_deuterons}.}
   	\label{fig_v1_pt}
\end{figure}

Figure \ref{fig_v1_pt} shows the directed flow as a function of transverse momentum for different rapidity bins.
Also as a function of $p_T$, the hard equation of state describes the experimental data best.
The difference between the two options to account for nuclei formation is clearly visible at low transverse momentum but vanishes at larger $p_T$.
The importance of clustering in the low momentum region can be explained by the large phase-space density that makes the formation of nuclei more likely.
Compared to the data, clustering gives the best description and  produces a small kink in the low $p_T$ region, while the explicit deuterons give a reasonable description of the data and the curve is more smooth but does not follow the data as closely.

\begin{figure}
	\centering
	\includegraphics[width=0.45\textwidth]{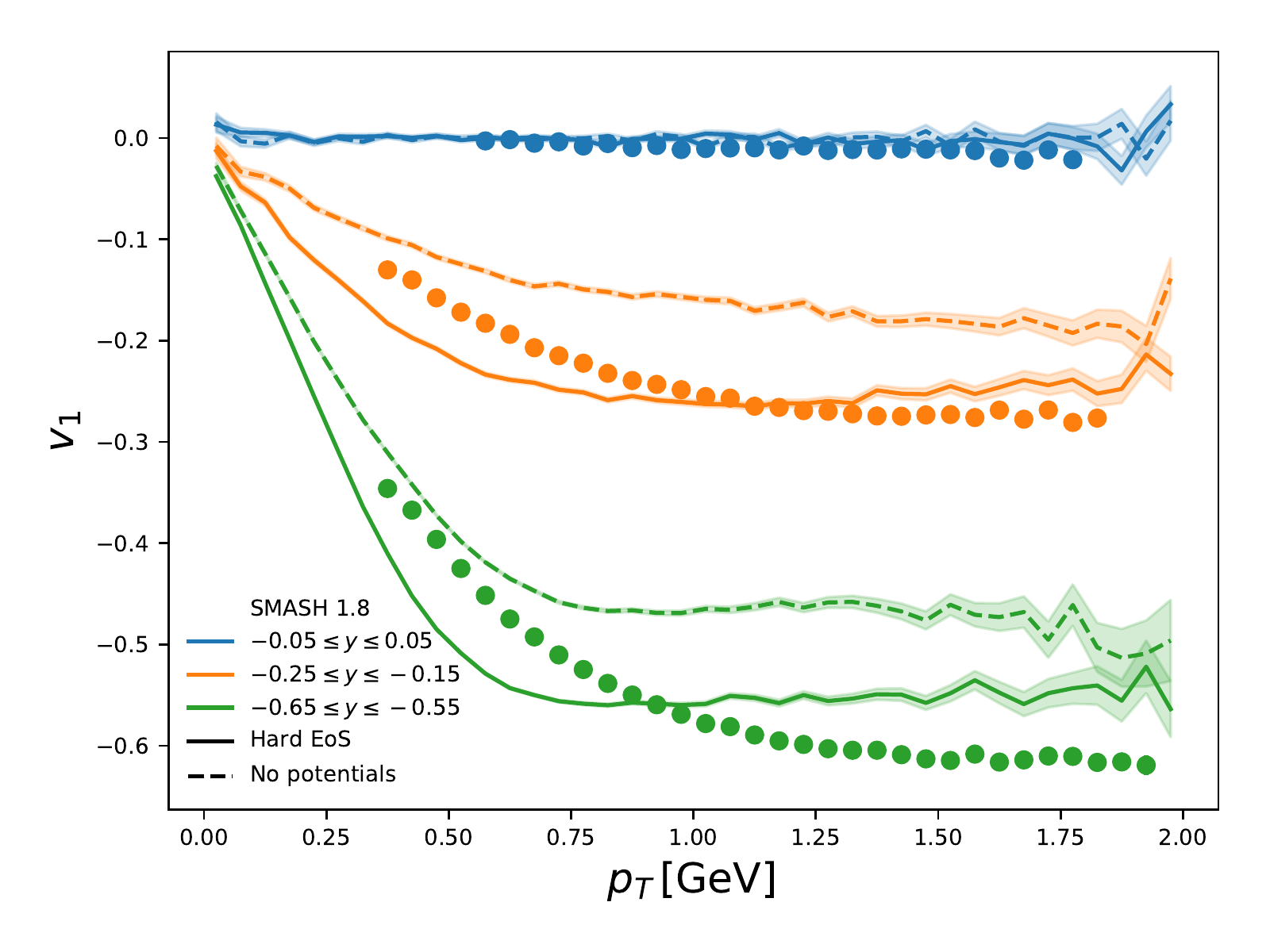}
	\caption{Directed flow of deuterons as a function of transverse momentum in 20\%-30\% central gold-gold collisions at $E_\mathrm{kin} = 1.23A\,\mathrm{GeV}$ for different rapidity bins compared to experimental data points \cite{Kardan:2018hna}. A hard equation of state was employed here and deuterons were dynamically treated as particles in this calculation.}
	\label{fig_deuteron_v1}
\end{figure}

From the calculation where deuterons are explicitly propagated, the directed flow of deuterons themselves is extracted. 
The results shown in Figure \ref{fig_deuteron_v1} compare the deuteron flow with and without nuclear potentials.
For the calculation with potentials, the hard equation of state is used. 

Considering that the model was originally designed for high energy collisions where the composition of the system is very different and nuclear potentials are negligible, it is not obvious that the directed flow of deuterons can be described. A surprisingly good agreement with the experimental data is observed when the potentials are included.

To conclude the findings from the comparison to the measured $v_1$, the hard equation of state is clearly favored in this setup.
Overall a good agreement with the data is observed, where the clustering setup follows a bit closer the transverse momentum dependent $v_1$ of nucleons, while the deuteron flow is matched extremely well in the calculation with explicit deuteron formation.  

\section{Elliptic flow of protons and deuterons}
\label{sec_elliptic_flow}
\begin{figure}
	\centering
	\includegraphics[width=0.45\textwidth]{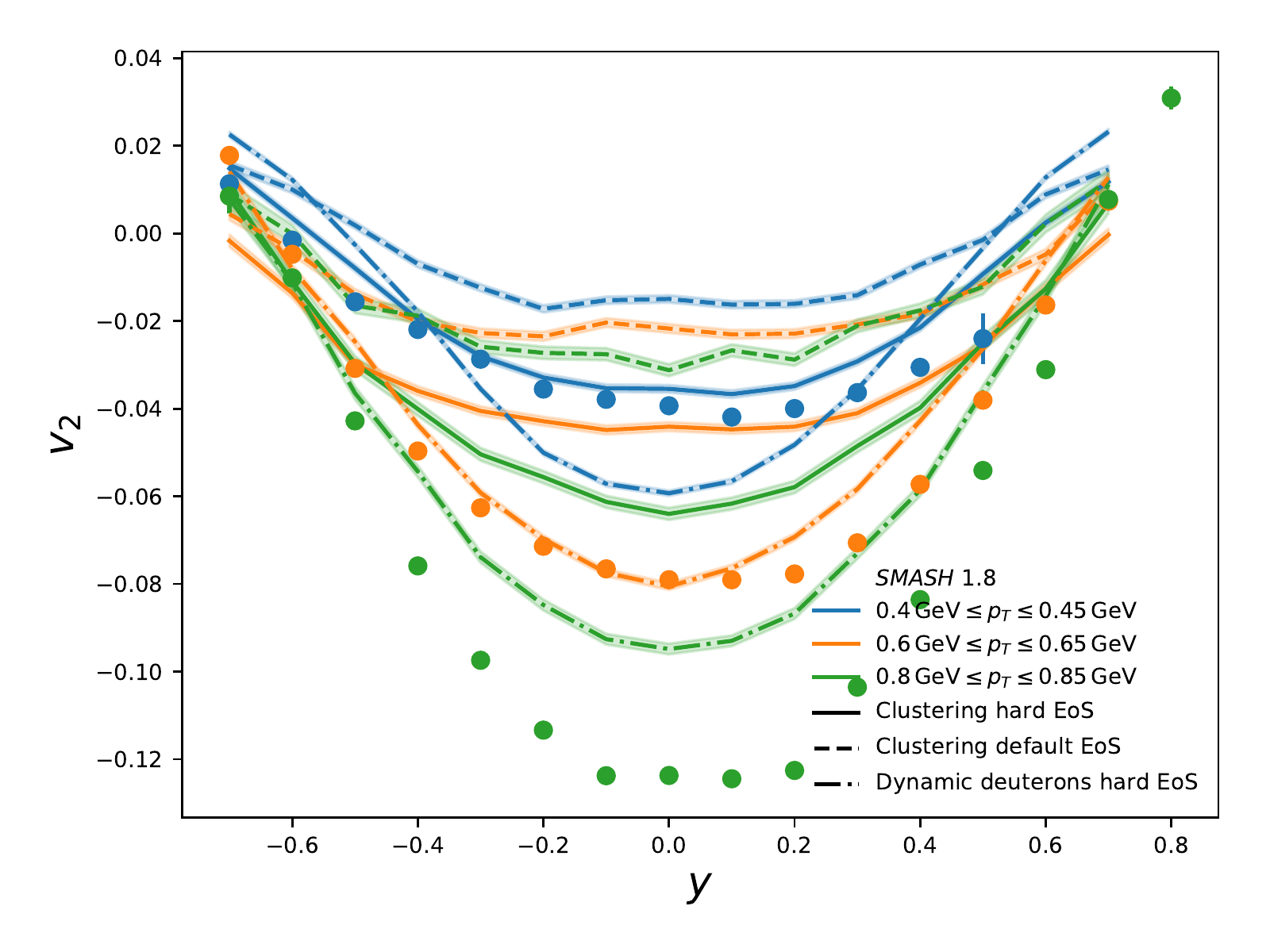}
	\caption{Elliptic flow of protons as a function of rapidity in 20\%-30\% central gold-gold collisions at $E_\mathrm{kin} = 1.23A\,\mathrm{GeV}$ for different $p_T$ bins compared to experimental data points \cite{Kardan:2018hna}. The full lines are obtained with a hard equation of state while for the dashed lines the default equation of state is used.}
	\label{fig_v2_y}
\end{figure}

Continuing with the second order flow harmonic, Figure \ref{fig_v2_y} shows the $v_2$ as a function of rapidity in semi central gold-gold collisions.
Again, two calculations with clustering and different equations of state and a calculation with explicit deuterons employing a hard equation of state are compared.
Naturally, the second order flow coefficient is more difficult to describe than the first order one.
In general, the magnitude of $v_2$ is too small at large transverse momenta.
However the flow in the intermediate $p_T$ region, where most nucleons are located, is comparable to the experimental data.
Same as for $v_1$, the hard equation of state produces a stronger flow signal and is therefore preferred by the data.
In addition to that, the magnitude of $v_2$ is larger when the deuterons are treated explicitly in the calculation.
That calculation agrees best with the data.
Similar to the observation for $v_1$, in the low $p_T$ region (here the $0.4\,\mathrm{GeV} < p_T <0.45\,\mathrm{GeV}$ bin) the results are very sensitive to how the formation of light nuclei is taken into account.
  
\begin{figure}
	\centering
	\includegraphics[width=0.45\textwidth]{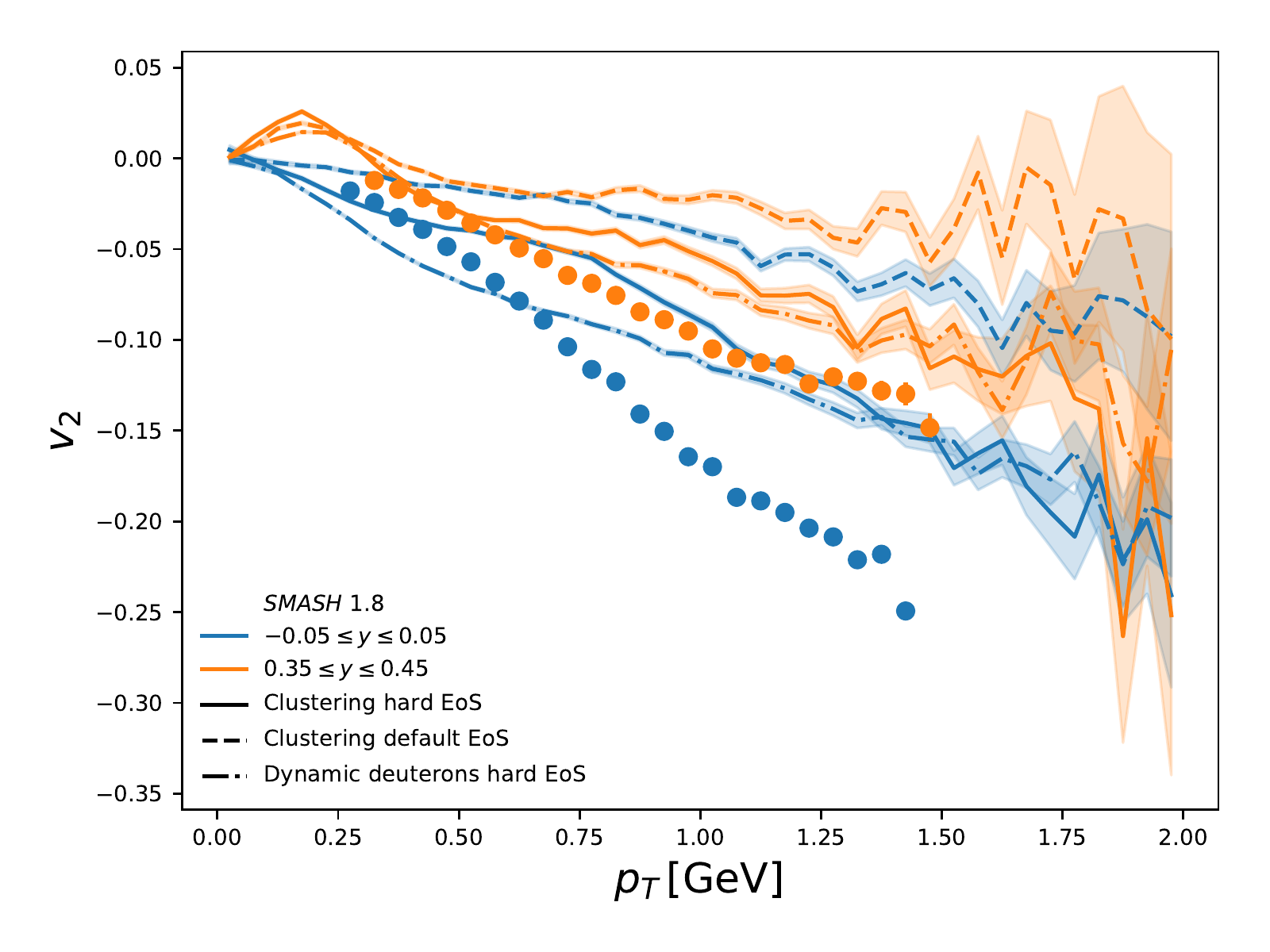}
   	\caption{Elliptic flow of protons as a function of transverse momentum in 20\%-30\% central gold-gold collisions at $E_\mathrm{kin} = 1.23A\,\mathrm{GeV}$ for different rapidity bins compared to experimental data points \cite{Kardan:2018hna}. The full lines are obtained with a hard equation of state while for the dashed lines the default equation of state is used.}
   	\label{fig_v2_pt}
\end{figure}

The elliptic flow of nucleons as a function of the transverse momentum is shown in Figure \ref{fig_v2_pt}.
Here it is again rather obvious that the elliptic flow is not well described, especially in the high $p_T$ region.
In the low $p_T$ region, the clustering performs better, while the calculation with explicit deuteron production works better at intermediate transverse momentum.
The difficulties in describing the $v_2$ at high $p_T$ possibly comes from the lack of momentum dependence of the potentials \cite{Welke:1988zz}.

\begin{figure}
	\centering
	\includegraphics[width=0.45\textwidth]{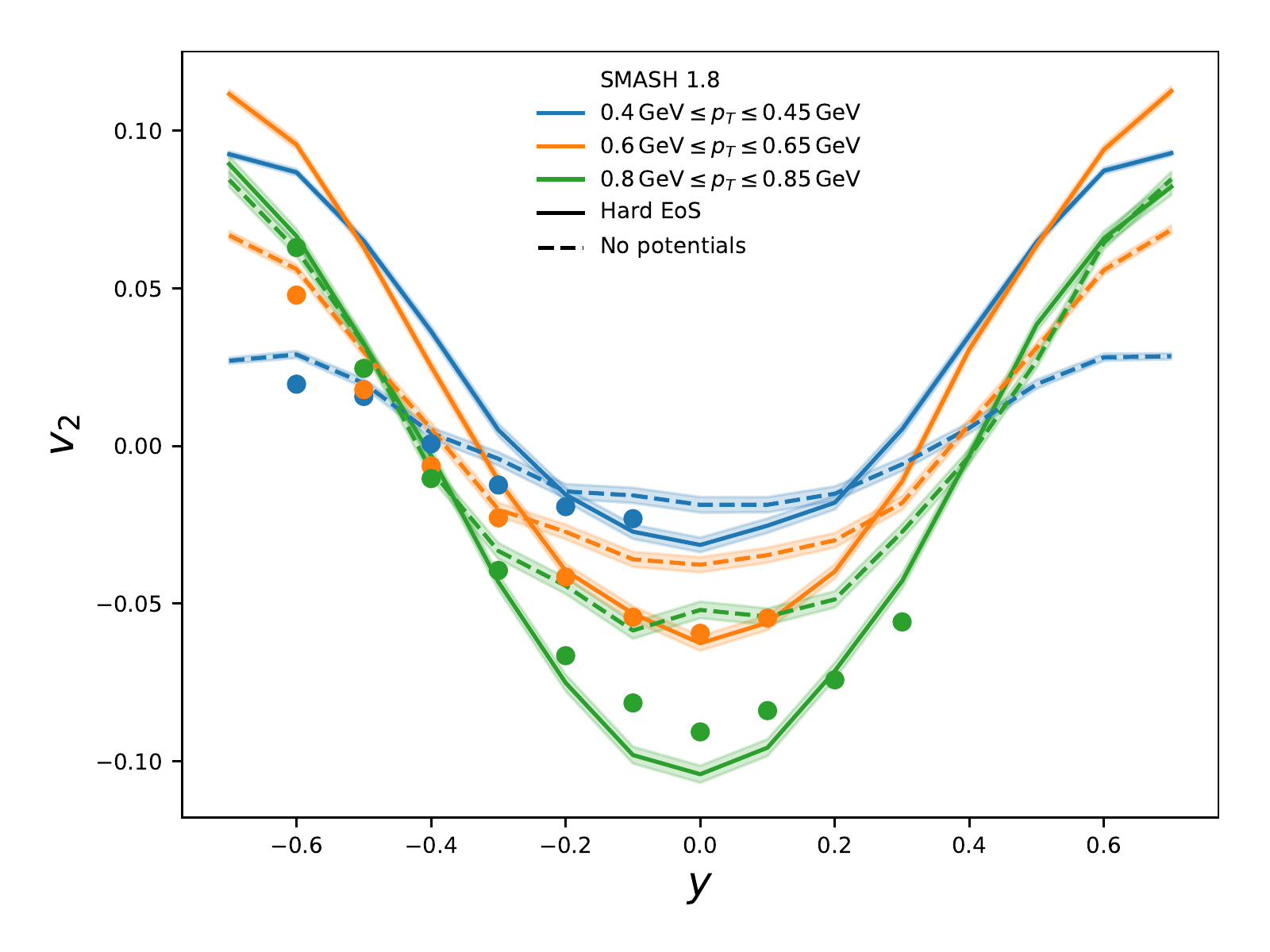}
	\caption{Elliptic flow of deuterons as a function of rapidity in 20\%-30\% central gold-gold collisions at $E_\mathrm{kin} = 1.23A\,\mathrm{GeV}$ for different transverse momentum bins compared to experimental data points \cite{Kardan:2018hna}. A hard equation of state was employed here and deuterons were dynamically treated as particles in this calculation.}
	\label{fig_deuteron_v2_y}
\end{figure}

\begin{figure}
	\centering
	\includegraphics[width=0.45\textwidth]{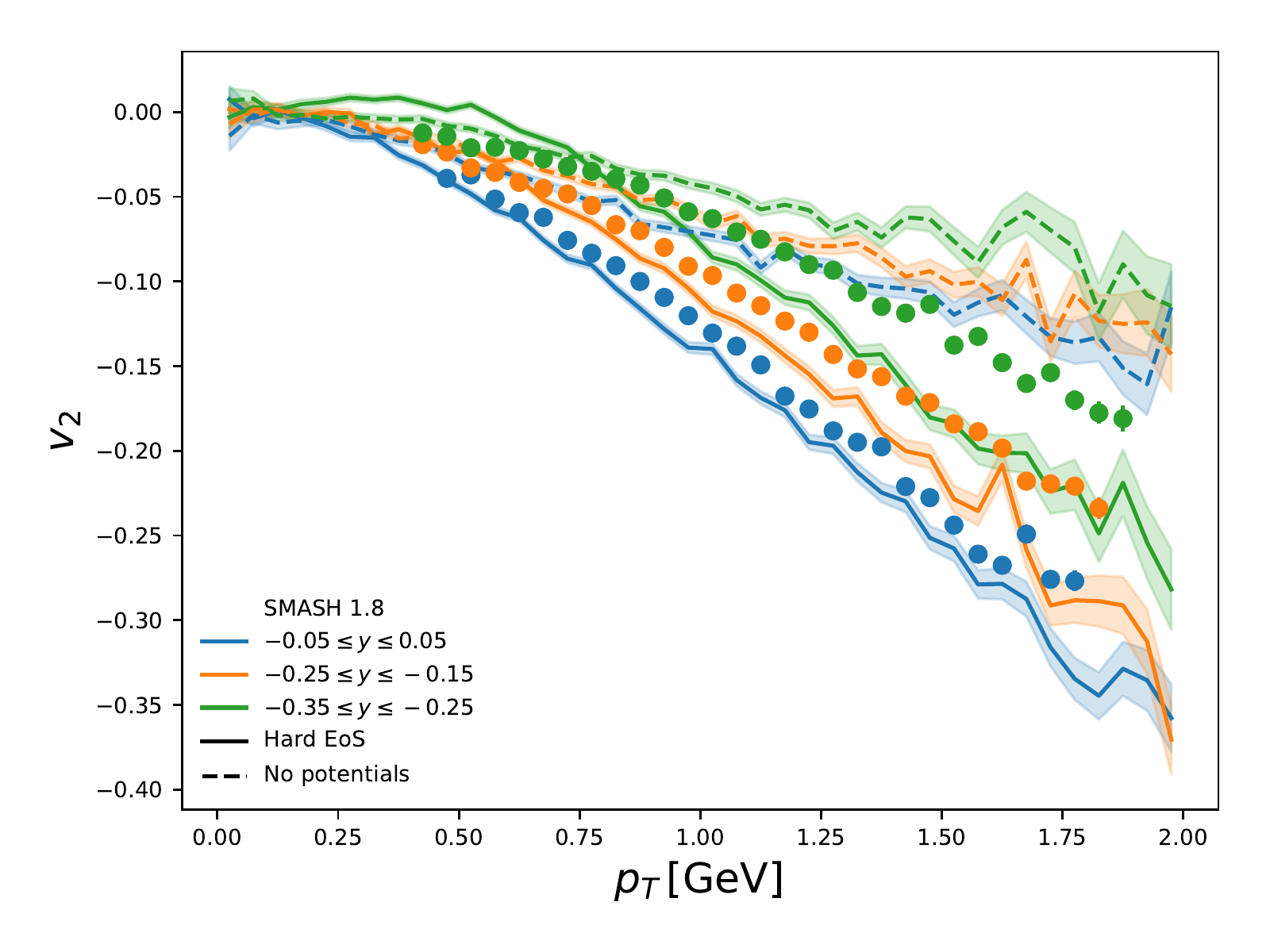}
	\caption{Elliptic flow of deuterons as a function of transverse momentum in 20\%-30\% central gold-gold collisions at $E_\mathrm{kin} = 1.23A\,\mathrm{GeV}$ for different rapidity bins compared to experimental data points \cite{Kardan:2018hna}. A hard equation of state was employed here and deuterons were dynamically treated as particles in this calculation.}
	\label{fig_deuteron_v2_pt}
\end{figure}

Figure \ref{fig_deuteron_v2_y} shows the elliptic flow of deuterons for the calculation, where they are treated as active degrees of freedom.
Same as for the directed flow, the calculation matches the data pretty well.
Also looking at the $p_T$ differential elliptic flow of deuterons in Figure \ref{fig_deuteron_v2_pt}, the agreement with the data is good.
At forward rapidity the calculation is slightly off but at mid-rapidity the data is described almost perfectly.

To conclude this section, the elliptic flow is observed to be very sensitive on how the light nuclei formation is taken into account, which leads to some uncertainty for the extraction of the EoS.
The elliptic flow of nucleons at large transverse momenta is underestimated in the SMASH calculations but the best agreement is found with the hard equation of state and treating the deuterons as active degrees of freedom.

\section{Evolution of flow coefficients}
\label{sec_evolution}
In this section we focus on the evolution of flow with time to see when the anisotropy is developed and what the most important stages are.
We concentrate on the setting that worked best in comparison to experimental data, namely the calculation with explicit deuteron formation and a hard equation of state. 

\begin{figure}
	\centering
	\includegraphics[width=0.45\textwidth]{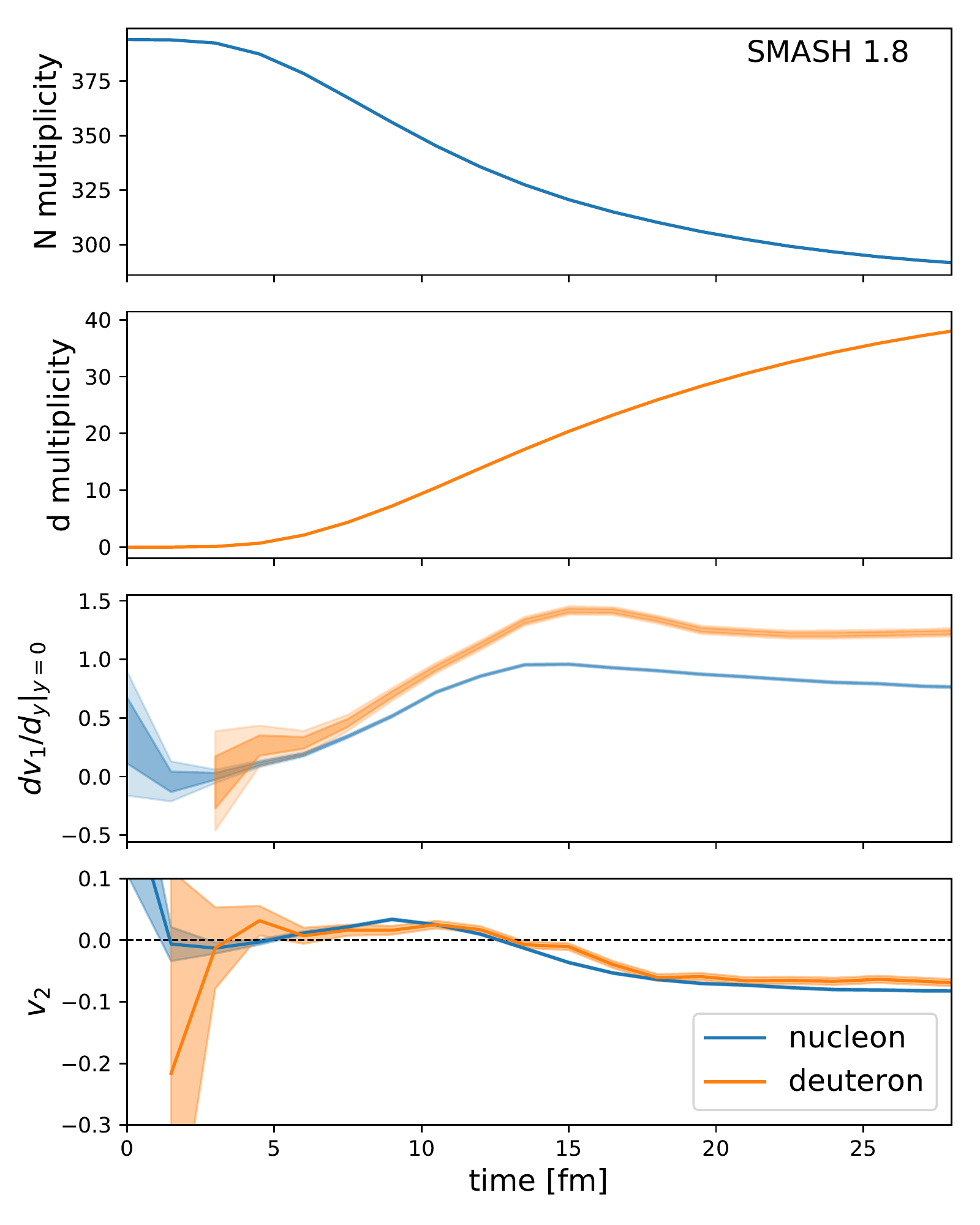}
	\caption{Nucleon and deuteron multiplicities, slope of the first order flow coefficient at mid-rapidity and $v_2$ of nucleons and deuterons as a function of time in 20\% - 30\% central gold-gold collisions at $E_\mathrm{kin}=1.23A\,\mathrm{GeV}$. The flow coefficients are evaluated for particles that satisfy $0.75\, \mathrm{GeV} < p_T < 0.8\,\mathrm{GeV}$ while the number of particles refers to the total multiplicity.}
	\label{fig_flow_of_time}
\end{figure}

The first two panels of Figure \ref{fig_flow_of_time} show the multiplicity of nucleons and deuterons respectively as a function of time.
The initially present nucleons from the two gold nuclei start interacting, forming resonances and deuterons, so that their number decreases. If no deuterons are present in the calculation, the number of nucleons increases again towards the end of the calculation when all resonances decay.
The number of deuterons rises throughout the evolution of the heavy ion collision.
In the third panel one can see the slope of the directed flow at mid-rapidity as a proxy for the magnitude of $v_1$ of nucleons and deuterons.
To evaluate the slope, at each point in time the $v_1(y)$ is fit with a function $v_1(y)=ay+by^3$ where the cubic term is added to take the curvature observed in Figure \ref{fig_v1_y} into account.
The fitting is performed using Bayesian parameter estimation with Markov chain Monte Carlo sampling. The inner band gives the 68\% and the outer band the 95\% credible interval.
At early times the uncertainty is very large since not enough particles have interacted yet and are not located in the relevant phase-space region and deuterons are not yet produced.
Afterwards the flow starts building up from zero to a maximum after $\approx 15$ fm.
This is the time the nuclei take to pass through each other.
Afterwards one observes for both nucleons and deuterons that the flow signal slightly weakens before saturating but the effect is more pronounced for deuterons.

The final panel of Figure \ref{fig_flow_of_time} shows a comparison of the elliptic flow at mid-rapidity between nucleons and deuterons.
Once enough particles are produced and located in the investigated kinematic region the elliptic flow of nucleons and deuterons look almost exactly the same.
This is not intuitively clear since no mass number scaling is applied here.
Looking at the evolution of $v_2$ in more detail one can see that at first a positive elliptic flow signal starts to build up but soon the $v_2$ drops below zero where it stays until the end of the evolution.
Positive elliptic flow is typically associated with pressure gradients in the initial state.
At low collision energies a competing effect is the squeeze out that results in a negative $v_2$ signal when the slow spectator nuclei are blocking the path and push particles out of the reaction plane \cite{Gutbrod:1988hh}.
Both effects contribute to the observed elliptic flow.

It is interesting to note that especially for nucleons the $v_2$ shows a very different behavior as a function of time than $v_1$. 
The difficulties in describing the elliptic flow and directed flow with the same parameter set might arise from the two being sensitive to different stages of the evolution.

\section{Higher flow coefficients}
\label{sec_higher_coefficients}
\subsection{Scalar product $v_3$}
In this section we would like to quantify the triangularity of heavy ion collisions to extend the excitation function from \cite{Karpenko:2015xea} down to $1.23A$ GeV. 
Unlike for the first and second harmonic, the $v_3$ signal does not simply emerge from the geometry of the colliding nuclei but rather by fluctuations in the overlap region.
To really quantify the triangularity  of an event it is therefore not sufficient to calculate the flow coefficient with respect to the reaction plane.
Hence for evaluating the triangular flow in this section the scalar product method \cite{Adler:2002pu} is employed because even though the HADES collaboration only measures $v_3(\Psi_2)$ we would like to explore, if there is a finite triangular flow in our calculations at low beam energies (see Appendix \ref{app_rpv34} for higher order flow coefficients with respect to the reaction plane).

In the scalar product method the $n$th flow coefficient is calculated as follows:
\begin{equation}
	v_n=\frac{\left\langle\vec{u} \cdot \frac{\vec{Q}_n}{N}\right\rangle}{\sqrt{\left\langle\frac{\vec{Q}_n^A}{N_A}\cdot\frac{\vec{Q}_n^B}{N_B}\right\rangle_E}}\,,
	\label{eq_vn_scalar_product}
\end{equation} 
where $\vec{u}$ is the momentum unit vector of a particle in the transverse plane, $\langle...\rangle$ denotes the average over all particles of interest while $\langle...\rangle_E$ is an average over events.
$N$ is the total number of particles and $N_A$ and $N_B$ are the number of particles in the sub-events A and B respectively. The same labeling applies for the flow-vectors $\vec{Q}^A_n$, $\vec{Q}^B_n$ and $\vec{Q}_n$ defined as
\begin{equation}
	\vec{Q}_n = 
	\begin{pmatrix}
	\sum_{i} w_i\cos(n\phi_i)\\ 
	\sum_{i} w_i\sin(n\phi_i)
	\end{pmatrix} \,,
\end{equation}
where $\phi_i$ is the azimuthal angle of particle $i$, $w_i$ is the weight which is for the $v_3$ calculation $p_T^3$.
The sum runs over the particles in the (sub-)event but for the scalar product in the numerator of Equation \ref{eq_vn_scalar_product} the particle of interest needs to be excluded from the flow vector calculation to avoid auto correlations.
Each event is divided into sub-events according to the pseudo-rapidity $\eta$ of the particles. One sub-event contains all particles with $\eta>0.1$ while the other one has particles that satisfy $\eta<-0.1$.

\begin{figure}
	\centering
	\includegraphics[width=0.45\textwidth]{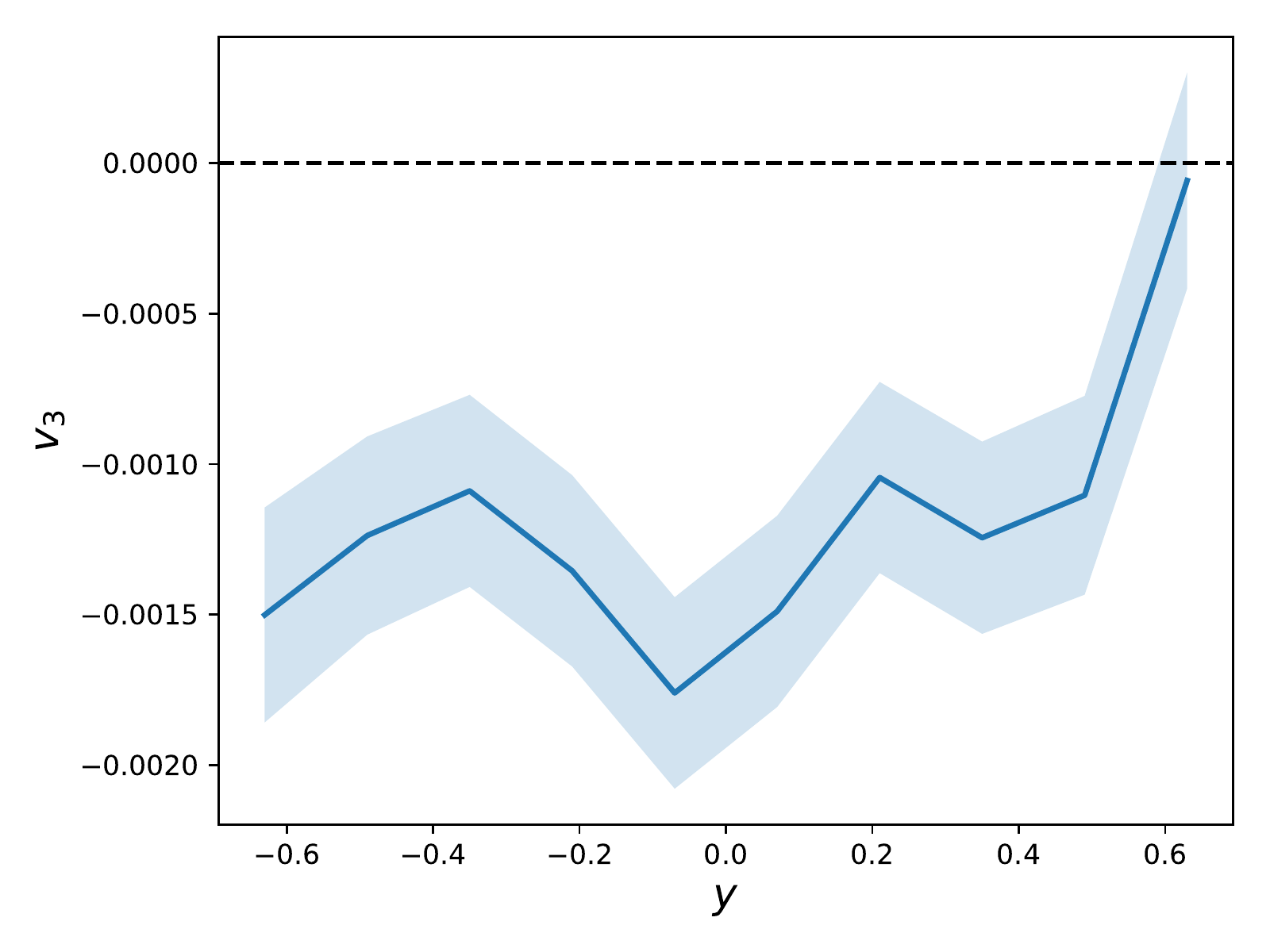}
	\caption{Triangular flow of nucleons as a function of rapidity in 20\%-30\% central gold-gold collisions. The flow coefficient is obtained using the scalar product method.}
	\label{fig_sp_v3}
\end{figure}

Applying the scalar product method to extract the triangular flow of nucleons one obtains the results shown in Figure \ref{fig_sp_v3}.
Looking at the scale of the axis the triangular flow signal in collisions at low energies almost vanishes.
That agrees with the findings from \cite{Karpenko:2015xea} where the triangular flow signal vanishes at low collision energies.
With the same cuts ($0.2\,\mathrm{GeV}<p_T< 2.0\,\mathrm{GeV}$ and $|\eta|< 1$) applied as in \cite{Karpenko:2015xea} we obtain for nucleons and deuterons $v_3=0.00081\pm 0.00006$ and $v_3=-0.0014\pm0.0002$ respectively.

\subsection{Quadrangular flow}
\begin{figure}
	\centering
	\includegraphics[width=0.45\textwidth]{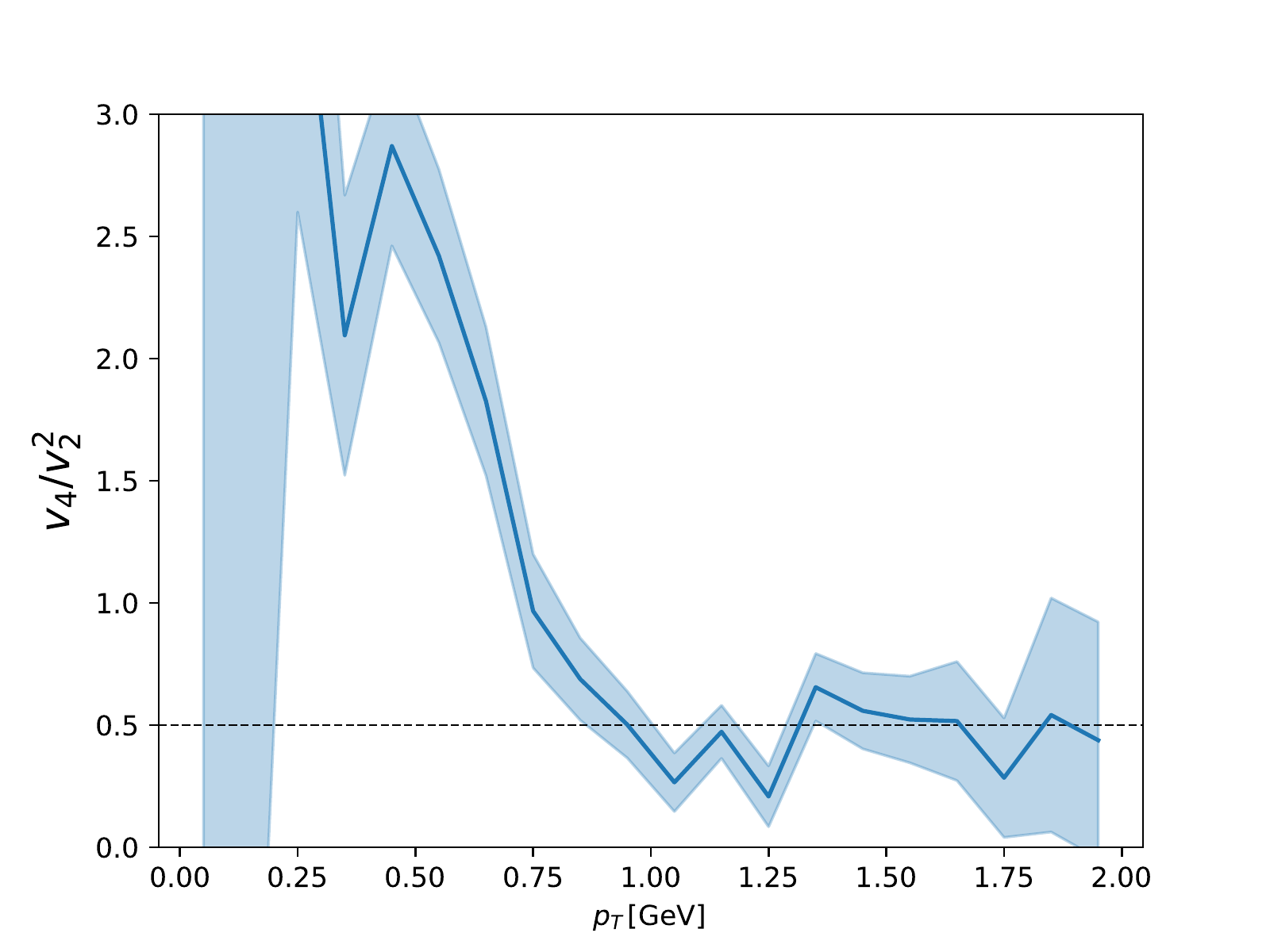}
	\caption{Quadrangular flow of nucleons divided by the squared elliptic flow at mid-rapidity as a function of transverse momentum in 20\% - 30\% central gold-gold collisions at $E_\mathrm{kin}=1.23A\,\mathrm{GeV}$.}
	\label{fig_v4_over_v2}
\end{figure}

In this section we show the fourth order flow coefficient of nucleons divided by the squared elliptic flow because this quantity is suggested to be a probe of ideal hydrodynamic behavior of the system if $v_4/v_2^2\approx 0.5$ \cite{Borghini:2005kd}.
$v_4/v_2^2$ is calculated at mid-rapidity as a function of $p_T$ and the result is shown in Figure \ref{fig_v4_over_v2}.
One can see that $v_4/v_2^2$ is close to 0.5 in the $p_T$ region above 0.75 GeV.
In general the shape of the curve is similar to the ideal three dimensional hydro prediction at RHIC energies from \cite{Borghini:2005kd} so the pure transport calculation as presented in this work shows in terms of this observable a close to hydrodynamic behavior. This is a sign that in low energy collisions the response to the initial geometry within a transport approach is translated as efficiently to final state observables as expected from a hydrodynamic picture. 
	
\section{Conclusions and Outlook}
\label{sec_conclusions}
In this paper we compared the double differential directed and elliptic flow of protons and deuterons to experimental flow measurements in Au+Au collisions at $E_{\rm lab}= 1.23A$ GeV as provided by the HADES collaboration.
Different parameter sets for the Skyrme potential, that each correspond to a different equation of state, are employed.
In our calculation for the flow of both  nucleons and deuterons a hard equation of state is preferred by the data.
The overall agreement for the elliptic flow of nucleons is still not satisfactory.
Therefore more sophisticated potentials including a momentum dependence will be necessary to extract the equation of state from flow measurements.

The main focus of the current work is the formation of light nuclei in low energy heavy-ion collisions that cannot be neglected even considering just the flow of nucleons.
Two different ways of taking the formation of deuterons into account are implemented.
Within this model we observe that dynamically forming deuterons and propagating them as particles throughout the calculation performs better with respect to the HADES data than forming deuterons via coalescence in the final state of a heavy-ion collision.
In the setup with the dynamic treatment of deuterons the evolution of the multiplicity and flow of nucleons and deuterons over time is studied.
Here one can see that the flow of nucleons and deuterons show a very similar time dependence, while the directed and elliptic flow coefficients are sensitive to different stages in the evolution. 

Finally we show for the first time that the scalar product triangular flow of nucleons almost vanishes in collisions at low energy and confirm that $v_4/v_2^2\approx0.5$ outside the low-$p_T$ region within the SMASH transport approach.

In the future a more advanced description of the nuclear potentials \cite{Sorensen:2020ygf} will allow for a more detailed comparison to a much broader set of experimental measurememts.
That comparison can be performed more systematically using Bayesian methods to put reliable constraints on the equation of state.    
	
\begin{acknowledgments}

Discussions and provision of experimental data is acknowledged to Behruz Kardan and Christoph Blume. 
J. M. acknowledges funding by GSI F\&E program. 
Computational resources have been provided by the Center for Scientific Computing (CSC) at the Goethe University and the GSI GreenCube. M. Mayer acknowledges support by the Deutsche Forschungsgemeinschaft (DFG) through the grant CRC-TR 211 "Strong-interaction matter under extreme conditions". M. Mayer would also like to thank Mathilde Ziegler-Himmelreich for fruitful discussions.
\end{acknowledgments}

\appendix
\section{Reaction plane $v_3$ and $v_4$}
\label{app_rpv34}
\begin{figure}[h]
	\centering
	\includegraphics[width=0.5\textwidth]{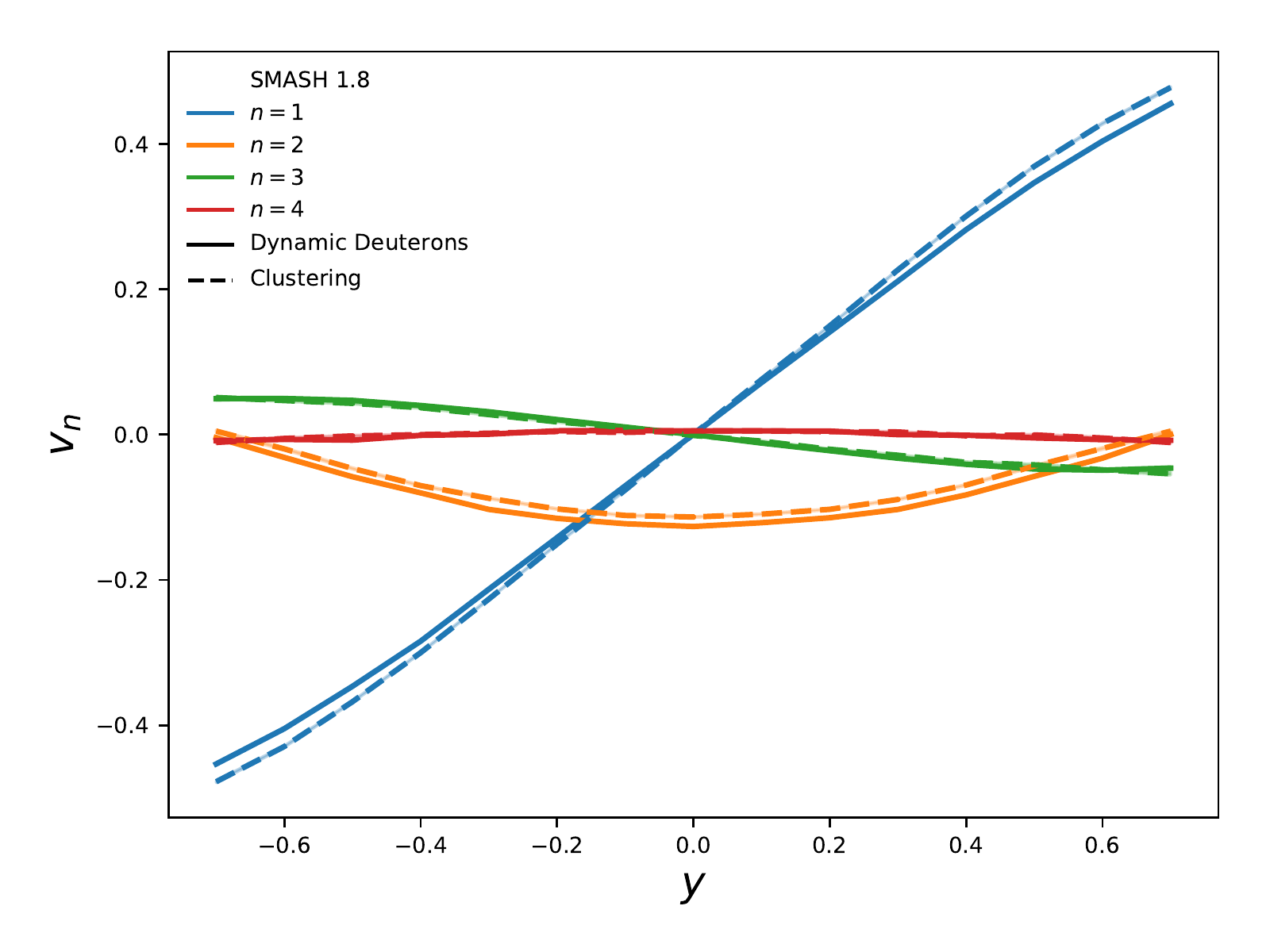}
	\caption{$v_1$ to $v_4$ of nucleons as a function of rapidity in 20\%-30\% central gold-gold collisions at $E_\mathrm{kin} = 1.23A\,\mathrm{GeV}$ for $1.0\,\mathrm{GeV} < p_T < 1.5\,\mathrm{GeV}$. All lines are obtained with a hard equation of state. Compared is a calculation with explicit deuteron formation and clustering.}
    \label{fig_rpv34_n}
\end{figure}

For completeness we provide the triangular and quadrangular flow with respect to the reaction plane in the same $p_T$ bin as shown in \cite{Adamczewski-Musch:2020iio}.
The hard equation of state is employed for all calculations in this appendix.
Figure \ref{fig_rpv34_n} shows the first four flow coefficients of nucleons as a function of rapidity.
Compared is a calculation with explicit deuteron formation as described in Section \ref{sec_deuterons} with final state clustering of nucleons \ref{sec_clustering}.

\begin{figure}[h]
	\centering
	\includegraphics[width=0.5\textwidth]{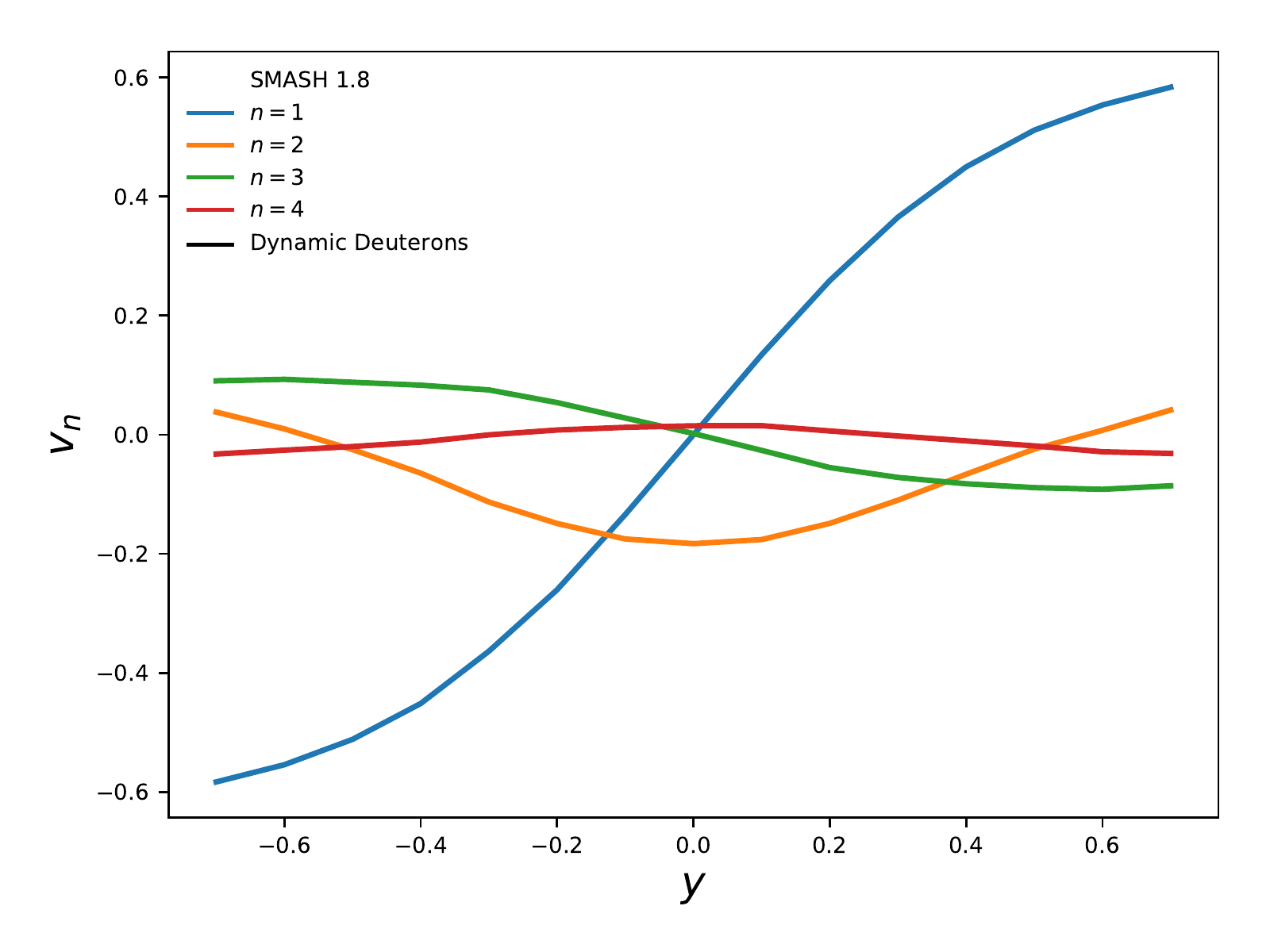}
	\caption{$v_1$ to $v_4$ of deuterons as a function of rapidity in 20\%-30\% central gold-gold collisions at $E_\mathrm{kin} = 1.23A\,\mathrm{GeV}$ for $1.0\,\mathrm{GeV} < p_T < 1.5\,\mathrm{GeV}$. All lines are obtained with a hard equation of state.}
	\label{fig_rpv34_d}
\end{figure}

We also provide the four lowest order flow coefficients calculation for deuterons in Figure \ref{fig_rpv34_d}. The deuterons are explicitly produced and propagated throughout the calculation.

\bibliography{flow_paper}

\end{document}